\documentclass[twocolumn,pre,aps,epsf]{revtex4-1}
\usepackage{graphicx}
\usepackage{amsmath,amssymb,amsbsy,amstext} 

\begin{document}

\title {Phase transition in a network model of social balance with Glauber dynamics}

\author{Rana Shojaei}
\author{Pouya Manshour}
\email{manshour@pgu.ac.ir}
\affiliation{Physics Department, Persian Gulf University, Bushehr 75169, Iran}
\author{Afshin Montakhab}
\email{montakhab@shirazu.ac.ir}
\affiliation{Physics Department, Shiraz University, Shiraz 71454, Iran}

\begin{abstract}
We study the evolution of a social network with friendly/enmity connections into a balanced state by introducing a dynamical model with an intrinsic randomness, similar to Glauber dynamics in statistical mechanics. We include the possibility of the tension promotion as well as the tension reduction in our model. Such a more realistic situation enables the system to escape from local minima in its energy landscape and thus to exit out of frozen imbalanced states, which are unwanted outcomes observed in previous models. On the other hand, in finite networks the dynamics takes the system into a balanced phase, if the randomness is lower than a critical value. For large networks, we also find a sharp phase transition at the initial positive link density of $\rho_0^*=1/2$, where the system transitions from a bipolar state into a paradise. This modifies the gradual phase transition at a nontrivial value of $\rho_0^*\simeq 0.65$, observed in recent studies.

\end{abstract}

\maketitle

\section{ Introduction }
\label{intro}
As Mark Buchanan discussed in his book, \textit{The Social Atom} \cite{Buchanan2007social}, we can think of people as elementary building blocks (the atoms) of the social world. The interactions between such interdependent elements lead to the emergence of macroscopic patterns such as cultures, wars, social classes, political parties, racial groups, etc. To understand such phenomena, one needs a new way of thinking, which borrows concepts from physics, in particular thermodynamics and statistical mechanics to study the macroscopic aspects of the human dynamics in social networks. Indeed, the evolution of our world is strongly ruled by social networks. In general, all political, economical, social or military conflicts occur in a social network, which includes a set of elements like countries, corporations or people that interact through different types of connections, such as friendship, hostility, political treaties, trade, or sharing ideas \cite{wasserman1994social,szell2010multi,
zheng2015social,kunegis2009slashdot,dodds2003information,
borgatti2003network,battiston2004statistical,powell2005network,montakhab2012low,manshour2014contagion}. 

Avoiding distress and conflict is a completely natural phenomenon in societies and interpersonal relationships, and almost all efforts are in the direction of tension reduction, if the individual nodes behave rationally \cite{Heider:1946,Cartwright:1956aa,harary1953notion}. In motivation psychology, Heider proposed a theory for attitude change, known as the balance theory \cite{Heider:1946}. By taking into account the relationship between three elements includes Person (P), and Other person (O) with an object (X), known as the POX pattern, he postulated that only balanced triads are stable. The POX is ``balanced" when P and O are friends, and they agree in their opinion of X. To reduce the stress in an imbalanced triad, the individuals change their opinions so that the triad becomes balanced. Empirical examples of Heider’s balance theory have been found in human and other animal societies \cite{Harary1961,Doreian1996,szell2010multi,Szell2010,Facchetti2011,Ilany2013}. Cartwright and Harary developed Heider model and showed that a complete signed graph with positive (agree) and negative (disagree) links is balanced, if and only if it can be decomposed into two fully positive subgraphs that are joined by negative links, called bipolar state \cite{Cartwright:1956aa,harary1953notion}.

For many years, authors only considered static signed networks. However, an important subject in the field of balance theory is the understanding of appropriate dynamics that can more accurately address the evolution of social networks, and explain how such a social balanced state emerges \cite{antal2006social,antal2005dynamics,kulakowski2005heider,radicchi2007social,marvel2011continuous,traag2013dynamical,kulakowski2007some,hummon2003some,altafini2012dynamics}. In such models one usually considers a complete network, i.e., all to all connections among nodes with dynamic signed links. Each link changes its status in order to reduce the local/global tension, if some conditions are satisfied. In a few works, continuous-valued links models have been investigated \cite{kulakowski2005heider,marvel2011continuous}, and have shown that the probability of reaching a balanced state in finite time tends to unity only for infinite system sizes. The influence of asymmetry in networks were also studied \cite{Traag2013}. Recently, the effect of memory on the evolution of the links has been investigated, which leads to a new glassy state in the networks \cite{hassanibesheli2017glassy}.

In an interesting work, Antal and colleagues introduced a dynamical model in complete networks with positive/negative links, called \textit{Constrained Triad Dynamics} (CTD) \cite{antal2006social,antal2005dynamics}. By definition, a triad is balanced if it has odd number of positive links. The update rule is as follows: A randomly chosen link is flipped, i.e., changes its sign, if the total number of imbalanced triads, $N_{imb}$, decreases. If $N_{imb}$ remains conserved, then the chosen link is flipped with probability $1/2$, and otherwise no changes in the link sign is allowed. This dynamics always takes the system into a more balanced situation. Two possible outcomes of such a dynamics are a balanced state or a \textit{jammed state}. A jammed state is an unwanted outcome in finite size, where the system is trapped into an imbalanced state, forever. They proved that the number of such jammed states greatly exceeds the number of balanced states, and that the probability of reaching them vanishes as the system size increases. By introducing an energy landscape, the properties of such jammed states have also been studied, extensively \cite{marvel2009energy,facchetti2012exploring}. Another result of CTD dynamics is a nontrivial gradual phase transition for the difference in sizes of the two final poles at $\rho_0\simeq 0.65$, where $\rho_0$ is the initial density of the positive links. They argued that this observed \textit{gradual} transition is not in agreement with analytical calculations.

As mentioned above, one of the fundamental assumptions in such dynamical models is the tendency to reduce the tension between elements of a social network, i.e., the update rules always take the system into a more balanced situation. But, this is not the case when we deal with the real world. For example, dissatisfaction, discomfort, profit, pride, anger, or generally speaking social anomalies, as well as random activity of each agent, are always present in social networks. In this article, we show, via detailed numerical and analytical calculations, that one can overcome the difficulties observed in previous models by introducing a more realistic dynamical model that takes into account the possibility of reduction as well as promotion of the tension among the social agents. We find some interesting results. Our dynamics is never trapped into a jammed state. In addition, the system undergoes a \textit{sharp} phase transition from a bipolar state into the paradise at $\rho_0=1/2$, for large networks.

\section{Model definition}
\label{method}

\begin{figure}[ht]
\begin{center}
\includegraphics[scale=.4]{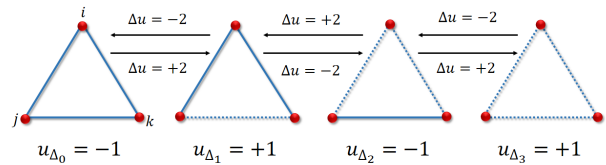}
\caption{Four distinct configurations of elementary units of the network. Solid lines show friendship links and dashed lines represent hostility links. The triads with energy  $-1$ ($+1$) are balanced (imbalanced).}
\label{fig1}
\end{center}
\end{figure}

In order to investigate how fluctuations in individual behaviors affect the balance theory, and also to overcome the observed shortcommings in previous studies, we introduce a dynamics as follows: We consider a fully connected network of size $N$, and use a symmetric conectivity matrix $s$, such that $s_{ij}=\pm1$. The positive sign represents friendship, and the negative one represents hostility between two arbitrary nodes $i$ and $j$. By definition of a triad to be of type $\Delta_k$ if it contains $k$ negative links, then $\Delta_0$ and $\Delta_2$ are balanced while $\Delta_1$ and $\Delta_3$ are imbalanced. Such conditions for balanced/imbalanced triads assert that a friend of my friend or an enemy of my enemy is my friend, and vice versa. For simplicity, one assumes that everyone knows everyone else, i.e., the dynamics occurs on a complete graph, which is appropriate for small real-world networks. As in \cite{antal2005dynamics,marvel2009energy}, the total energy of the system is defined as:
\begin{equation}
U=\frac{1}{N_{tri}}\sum_{i>j>k} u_{ijk}
\label{U}
\end{equation}
where $u_{ijk}=-s_{ij}s_{jk}s_{ki}$, and the normalization factor $N_{tri}=\binom{N}{3}$ is the total number of triads in the network, so we have $-1\leq U \leq 1$. By this definition, we have $u=-1$ and $+1$ for a balanced and an imbalanced triad, respectively. Also, $U=-1$ is the balance condition for the system, in which all the triads are balanced. At every time step, we flip a randomly chosen link with probability $p$, defined as
\begin{equation}
p=\frac{1}{1+e^{\beta \Delta U(t)}}
\label{p}
\end{equation}
where $\beta$ is a control parameter, and represents the inverse of the disorder in the system, which may be considered as the stochasticity in the individual behavior. Also, $\Delta U(t)$ indicates the change in the energy due to the link-flipping in every time step $t$. Fig.~\ref{fig1} shows all possible configurations of each triad due to an update. Such dynamics is similar to that of Glauber dynamics used in simulations of kinetic Ising model \cite{Glauber1963}. It is important to note that such dynamics provides a more realistic feature of creating or reducing tension at any given time while \textit{on average} reducing tension for positive finite $\beta$. Furthermore, since it allows for increase in the energy of the system, it could provide a natural mechanism to escape out of local minima, i.e., the jammed states.

We intend to study the dynamics of such a model, i.e., $\rho(t)$ and $U(t)$, for various initial configurations $\rho_0$ and randomness parameter $\beta$. We first provide mean-field analytical results for some special cases and then consider the model in more general conditions, numerically.

\section{Mean-field approach}
\label{sec:analytic}
Due to the large number of degrees of freedom in the system, finding exact time dependent equations for our dynamics is inaccessible. Thus, we try to find a mean-field approximation for the rate equations of our dynamics, using the notations used in \cite{antal2005dynamics}. It is appropriate to work with quantity $n_i$ which is the density of triads of type $\Delta_i$, i.e., $n_i=N_i/N_{tri}$, where $N_i$ is the number of such triads. With this definition, the number of positive links and the density of such links become $L_+=(3N_0+2N_1+N_2)/(N-2)$, and $\rho=L_+/L$, respectively, where $L=\binom{N}{2}$ is the total number of links and $L_{+}$ is the number of positive links. Thus, the  energy, $U$, and the density $\rho$ can be written as follows:
\begin{equation}
\begin{split}
\rho&=n_0+2n_1/3+n_2/3 \\
U&=-n_0+n_1-n_2+n_3
\end{split}
\label{rho_U}
\end{equation}
Another useful quantity is the density $n_i^+$ ($n_i^-$) of triads of type $\Delta_i$ that are connected to a positive (negative) link. The total number of positive links connected to triads of type $\Delta_i$ is $(3-i)N_i$, and $N_i^+=(3-i)N_i/L_+$ is the average number of such triads. Since, each link is connected to $N-2$ triads of any types, thus $n_i^+=N_i^+/(N-2)$. Similarly, one can obtain $n_i^-=N_i^-/(N-2)$ for a negative link. Finally, we can write
\begin{equation}
\begin{split}
n_i^+&=(3-i)n_i/(3n_0+2n_1+n_2) \\
n_i^-&=in_i/(n_1+2n_2+3n_3)
\end{split}
\label{n+-}
\end{equation}

Taking into account that $\rho$ is the probability of finding a positive link, the probability of flipping a \textit{positive} link is $\pi^+=p^+\rho$, with
\begin{equation}
p^+=\frac{1}{1+e^{\beta \Delta U_{+-}}}
\label{p_p}
\end{equation}
and of flipping a \textit{negative} link is $\pi^-=p^-(1-\rho)$, with
\begin{equation}
p^-=\frac{1}{1+e^{\beta \Delta U_{-+}}}
\label{p_m}
\end{equation}
where $\Delta U_{+-}$ and $\Delta U_{-+}$ are the energy difference due to the flipping a positive and a negative link, respectively. Thus, for an each update at step $n$, we have
\begin{equation}
L_+(n+1)-L_+(n)=-\pi^+ + \pi^-
\end{equation}
Since each time step equals $L$ updates, so one can simply find the rate equation for (average) $\rho$, as
\begin{equation}
\frac{d\rho}{dt}=-\pi^+ + \pi^-
\label{rho_rate}
\end{equation}

In each update, the energy difference due to the flipping of a positive link equals to $2(N_0^+-N_1^++N_2^+)/N_{tri}$, and similarly for the flipping of a negative link we have $2(-N_1^-+N_2^--N_3^-)/N_{tri}$. Thus we obtain
\begin{equation}
\begin{split}
U(n+1)-U(n)&=2\pi^+(N_0^+-N_1^++N_2^+)/N_{tri} \\
&+ 2\pi^-(-N_1^-+N_2^--N_3^-)/N_{tri}
\end{split}
\end{equation}
Therefore, we find the rate equation of the total energy as
\begin{equation}
\frac{dU}{dt}=\pi^+\Delta U_{+-} + \pi^-\Delta U_{-+}
\label{U_rate}
\end{equation}
where $\Delta U_{+-}=6(n_0^+-n_1^++n_2^+)$ and $\Delta U_{-+}=-6(n_1^--n_2^-+n_3^-)$. 
In a similar way one can also find the rate equations for all triad densities, $n_i$, as follows:
\begin{equation}
\begin{split}
\frac{dn_0}{dt}&=-3\pi^+ n_0^+ + 3\pi^-n_1^- \\
\frac{dn_1}{dt}&=-3\pi^+ n_1^+ - 3\pi^-n_1^- +3\pi^+ n_0^++3\pi^-n_2^- \\
\frac{dn_2}{dt}&=-3\pi^+ n_2^+ - 3\pi^-n_2^- +3\pi^+ n_1^++3\pi^-n_3^- \\
\frac{dn_3}{dt}&=-3\pi^- n_3^- + 3\pi^+ n_2^+
\end{split}
\label{dens_rate}
\end{equation}
where Eqs. (\ref{rho_rate}) and (\ref{U_rate}) can also be derived from Eqs. (\ref{rho_U}) and (\ref{dens_rate}).

For completely random flipping, $\beta\rightarrow 0$, we have $p^+=p^-=1/2$. Thus Eq. (\ref{rho_rate}) becomes $\frac{d\rho}{dt}=1/2-\rho$. Simply, we find that
\begin{equation}
\rho(t)=1/2+(\rho_0-1/2)e^{-t}
\label{rho_t}
\end{equation}
which $\rho_{\infty}=1/2$ independent of  $\rho_0$, as expected.
Due to the uncorrelated nature of the dynamics at $\beta=0$, the triad densities become $n_0=\rho^3$, $n_1=3\rho^2(1-\rho)$, $n_2=3\rho(1-\rho)^2$, and $n_3=(1-\rho)^3$. By substituting these densities into $\Delta U_{+-}$ and $\Delta U_{-+}$, using relations for $n_i^+$ and $n_i^-$, we find
\begin{equation}
\begin{split}
\Delta U_{+-}&=+24(\rho-1/2)^2 \\
\Delta U_{-+}&=-24(\rho-1/2)^2
\end{split}
\label{Delt_u}
\end{equation}
Thus, Eq. (\ref{U_rate}) becomes
\begin{equation}
\frac{dU}{dt}=24(\rho-1/2)^3
\end{equation}
One simply finds
\begin{equation}
U(t)=-8(\rho_0-1/2)^3e^{-3t}
\label{U_t}
\end{equation}
which shows that $U_{\infty}\rightarrow 0$, as expected for an uncorrelated network. Also, we find a relation between $U$ and $\rho$ for an uncorrelated network as:
\begin{equation}
U=-8(\rho-1/2)^3
\label{U_rho}
\end{equation}

For the case of large $\beta$, we first assume that the system remains uncorrelated during its early stages of the evolution, as observed from simulations for large networks (see the next section). Therefore, Eq. (\ref{Delt_u}) holds for initial time steps. Consequently, $p^-\rightarrow 1$, and $p^+\rightarrow 0$, and thus $\frac{d\rho}{dt}\simeq (1-\rho)$. We find the time behavior of $\rho$, as
\begin{equation}
\rho(t)=1+(\rho_0-1)e^{-t}
\label{rho_t1}
\end{equation}
and also for $U$ as
\begin{equation}
U(t)=-8(1/2+(\rho_0-1)e^{-t})^3
\label{U_t1}
\end{equation}
This shows that for large $t$, the dynamics takes the system into a paradise state, i.e., $\rho_{\infty}\rightarrow 1$ and $U_{\infty}\rightarrow -1$, independent of $\rho_0$.

On the other hand, one can find stationary solutions of the rate equations, for any arbitrary $\beta$. From Eqs. (\ref{rho_rate}), (\ref{U_rate}) and (\ref{dens_rate}), we find $\pi_+=\pi_-$,  $n_0^+=n_1^-$, $n_1^+=n_2^-$, $n_2^+=n_3^-$, and also $\Delta {U_{+-}}^*=-\Delta {U_{-+}}^*=+24(\rho_{\infty}-1/2)^2$. With a little algebra, one can obtain $U_{\infty}=-8(\rho_{\infty}-1/2)^3$, and  
\begin{equation}
\rho_{\infty}=\frac{{p^-}^*}{{p^-}^*+{p^+}^*}
\label{rho_inf}
\end{equation}
where 
\begin{equation}
\begin{split}
{p^+}^*=\frac{1}{1+e^{24\beta (\rho_{\infty}-1/2)^2)}}\\
{p^-}^*=\frac{1}{1+e^{-24\beta (\rho_{\infty}-1/2)^2)}}
\end{split}
\label{p_inf}
\end{equation}
Note that Eq. (\ref{rho_inf}) is a self-consistent equation for $\rho_{\infty}$. As it can be seen, the balanced ($U_{\infty}=-1$) stationary solution of the system only occurs for $\beta\rightarrow \infty$, with $\rho_{\infty}=1$. However, as we will demonstrate later, another balanced solution ($\rho_{\infty}=1/2$) also occurs for large $\beta$. We note here that these analytical relations are obtained by taking into account the mean-field approximation, where we assume the same dynamics for all triads of type $\Delta_i$, and the effects of correlations are averaged out. We will further show that for values of $\rho$ near $1/2$, correlation effects cannot be neglected, and our mean-filed results are not able to represent such regimes. Thus, our analytical results are not exact. In the next section, we will prove this claim and also discuss the dynamics near this point, in more details.

\section{Numerical Results}
\label{diss}

\begin{figure}[ht]
\begin{center}
\includegraphics[scale=.5]{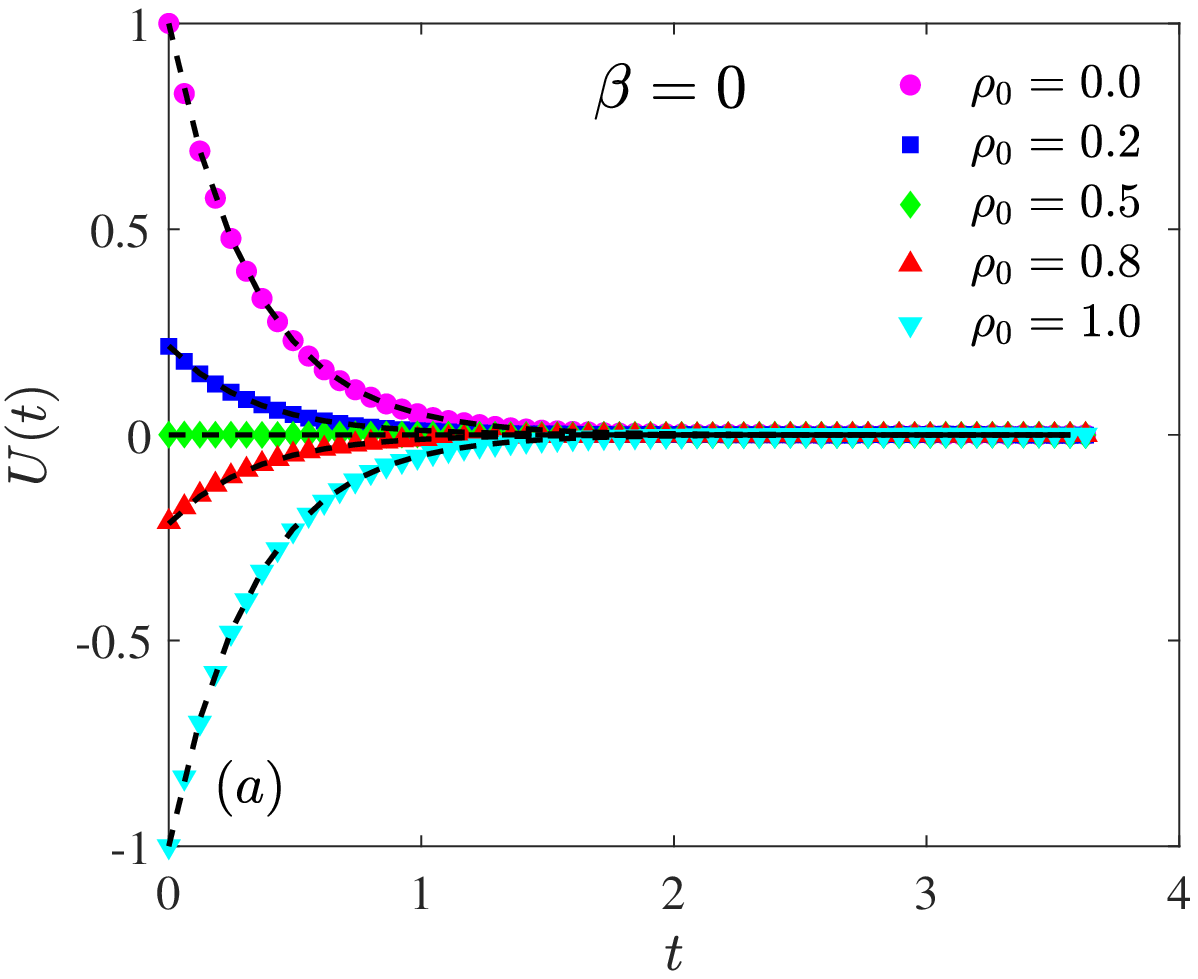}
\includegraphics[scale=.5]{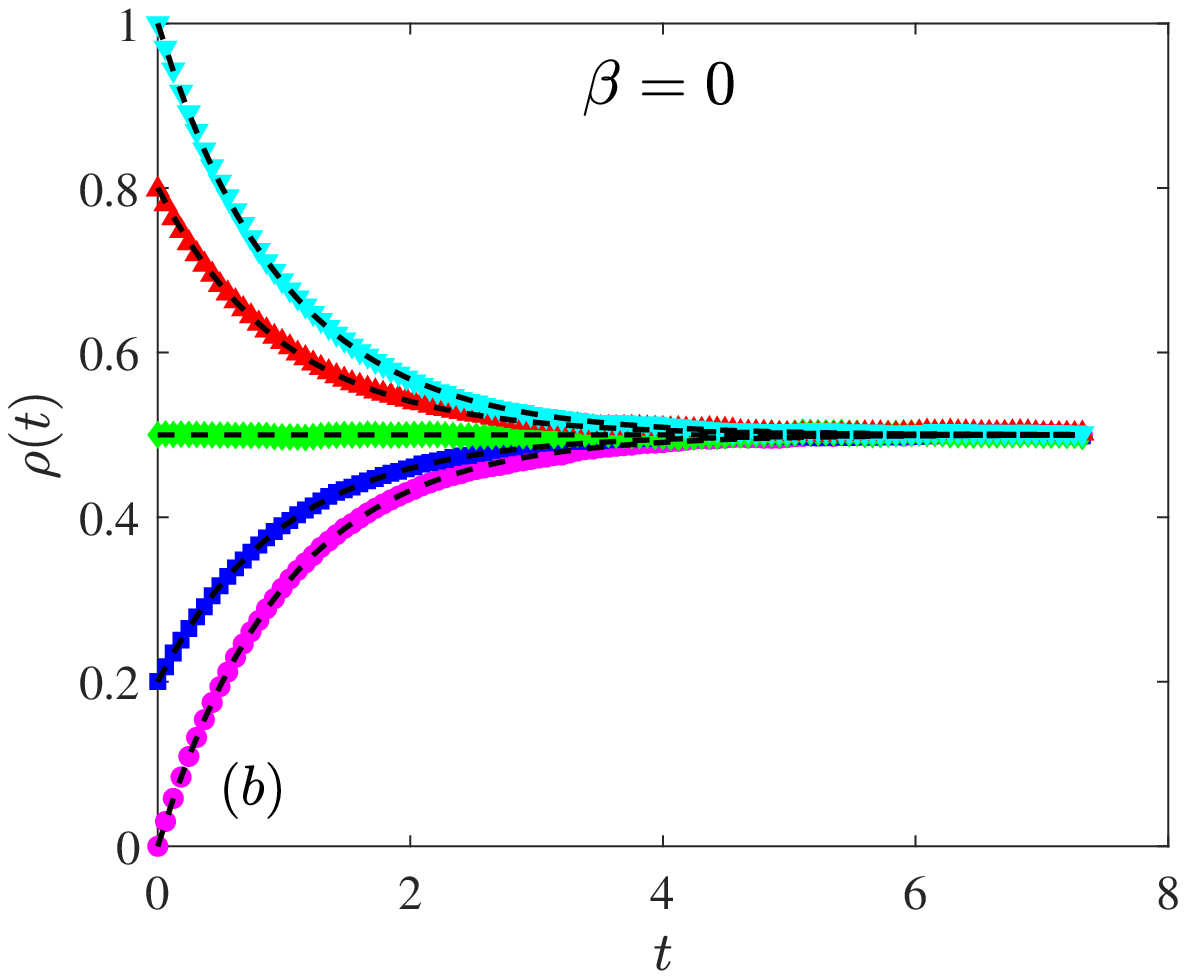}
\includegraphics[scale=.5]{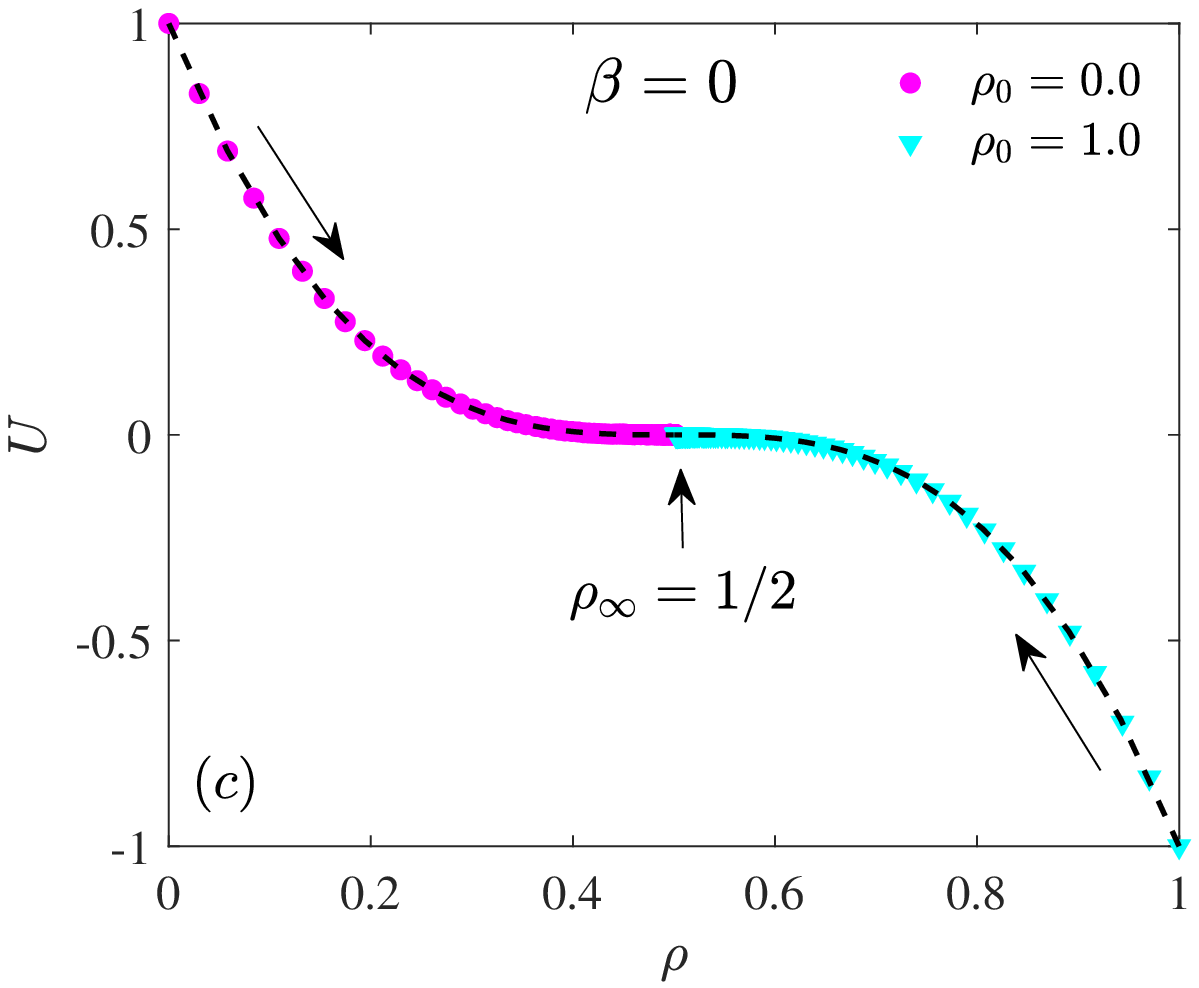}
\caption{The time evolution of (a) the total energy, $U$, and (b) the density of positive links, $\rho$, with different initial conditions for $\beta=0$. (c) The trajectory of this dynamics in $U-\rho$ space, for two initial conditions of $\rho_0=0$, and $1$. The network size for all plots is $N=128$. Note that the dashed lines in (a), (b), and (c) show our analytical solutions of Eqs. (\ref{rho_t}), (\ref{U_t}), and (\ref{U_rho}), respectively.}
\label{fig2}
\end{center}
\end{figure}

\begin{figure}[ht]
\begin{center}
\includegraphics[scale=.5]{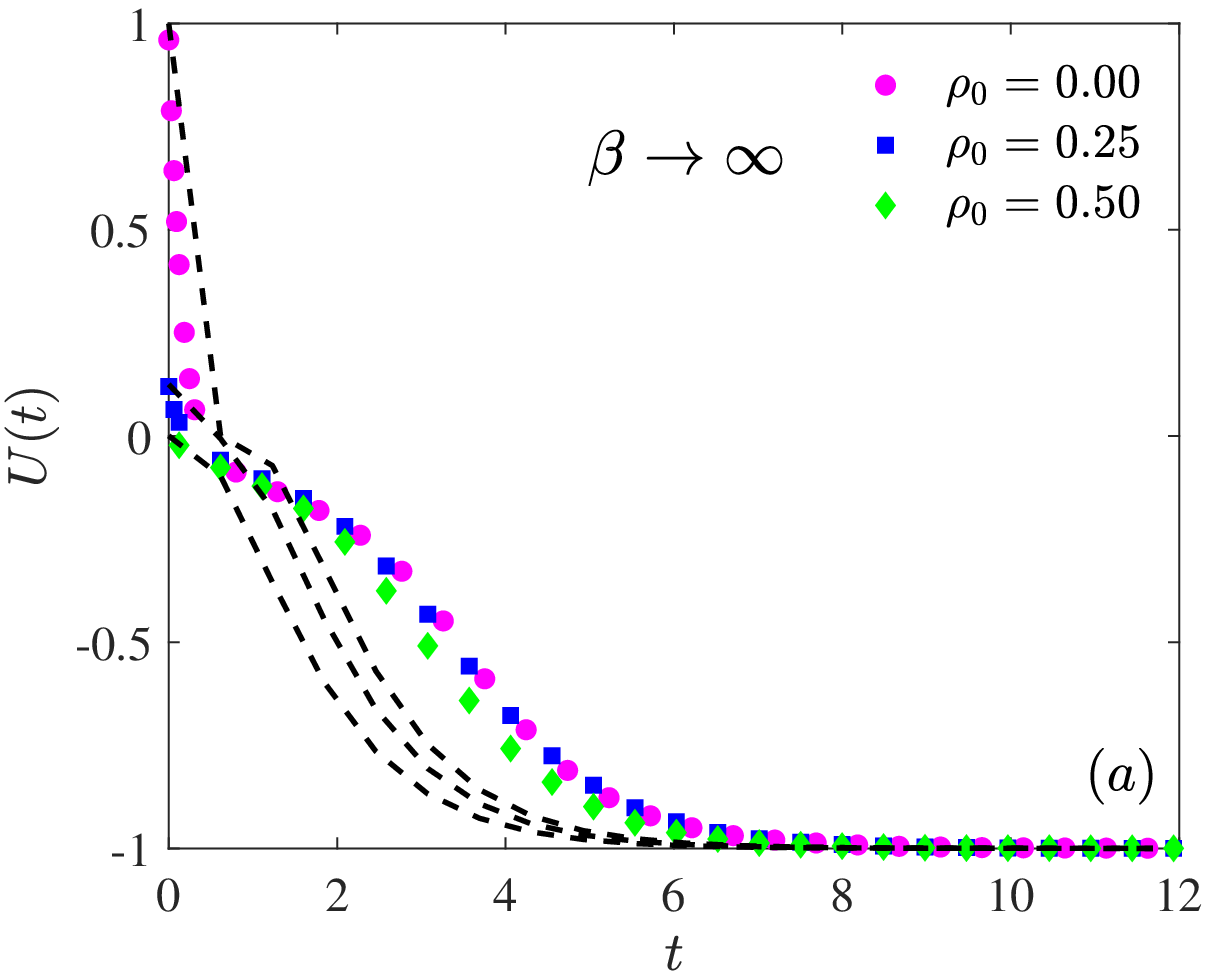}
\includegraphics[scale=.5]{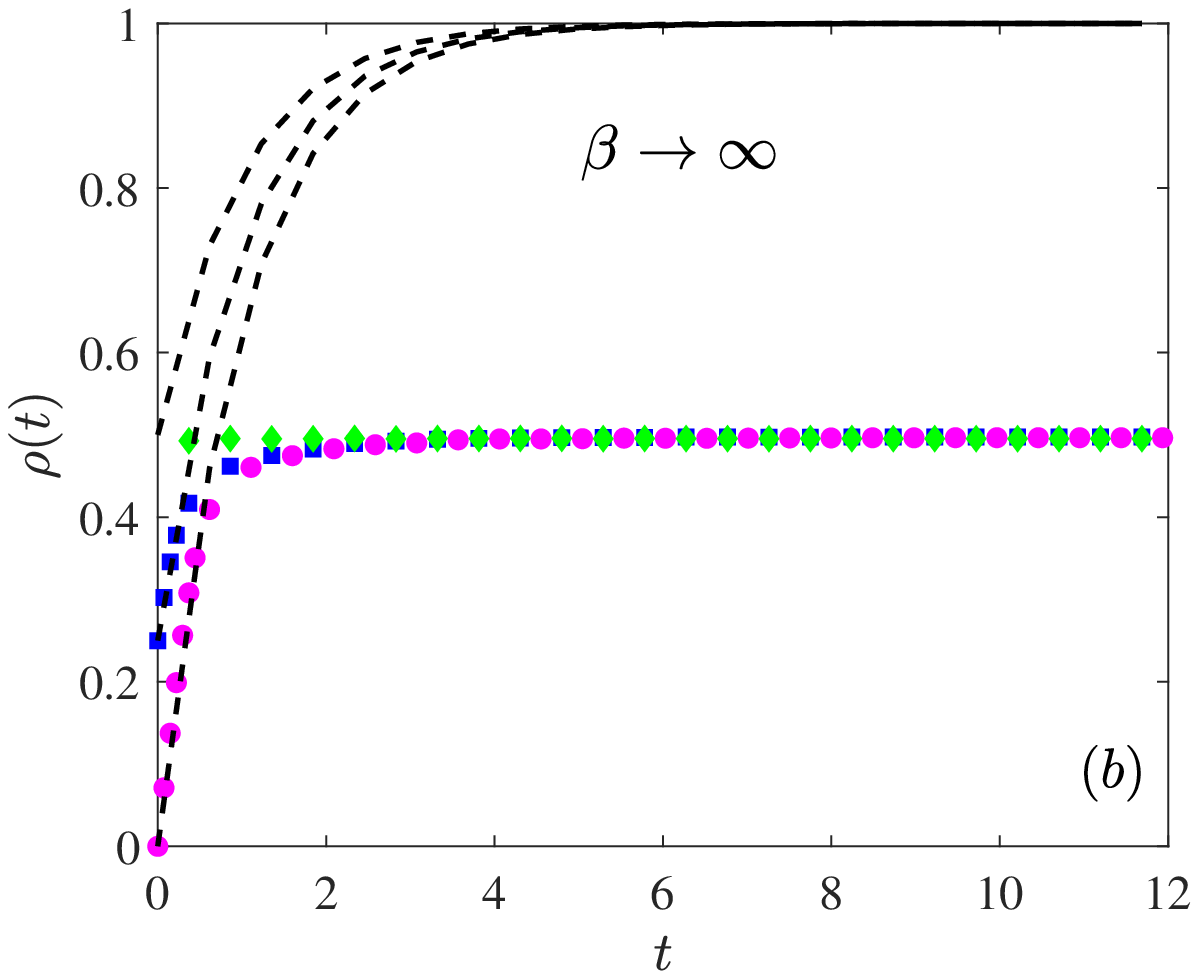}
\caption{The time evolution of (a) the total energy, $U$, and (b) the density of positive links, $\rho$, with different initial conditions of $\rho_0\leq 1/2$ for $\beta\rightarrow\infty$. The dashed lines show our analytical solutions, i.e., Eqs. (\ref{rho_t1}) and (\ref{U_t1}), which are not in agreement with the simulations. The network size for all plots is $N=128$.}
\label{fig3}
\end{center}
\end{figure}

\begin{figure}[ht]
\begin{center}
\includegraphics[scale=.5]{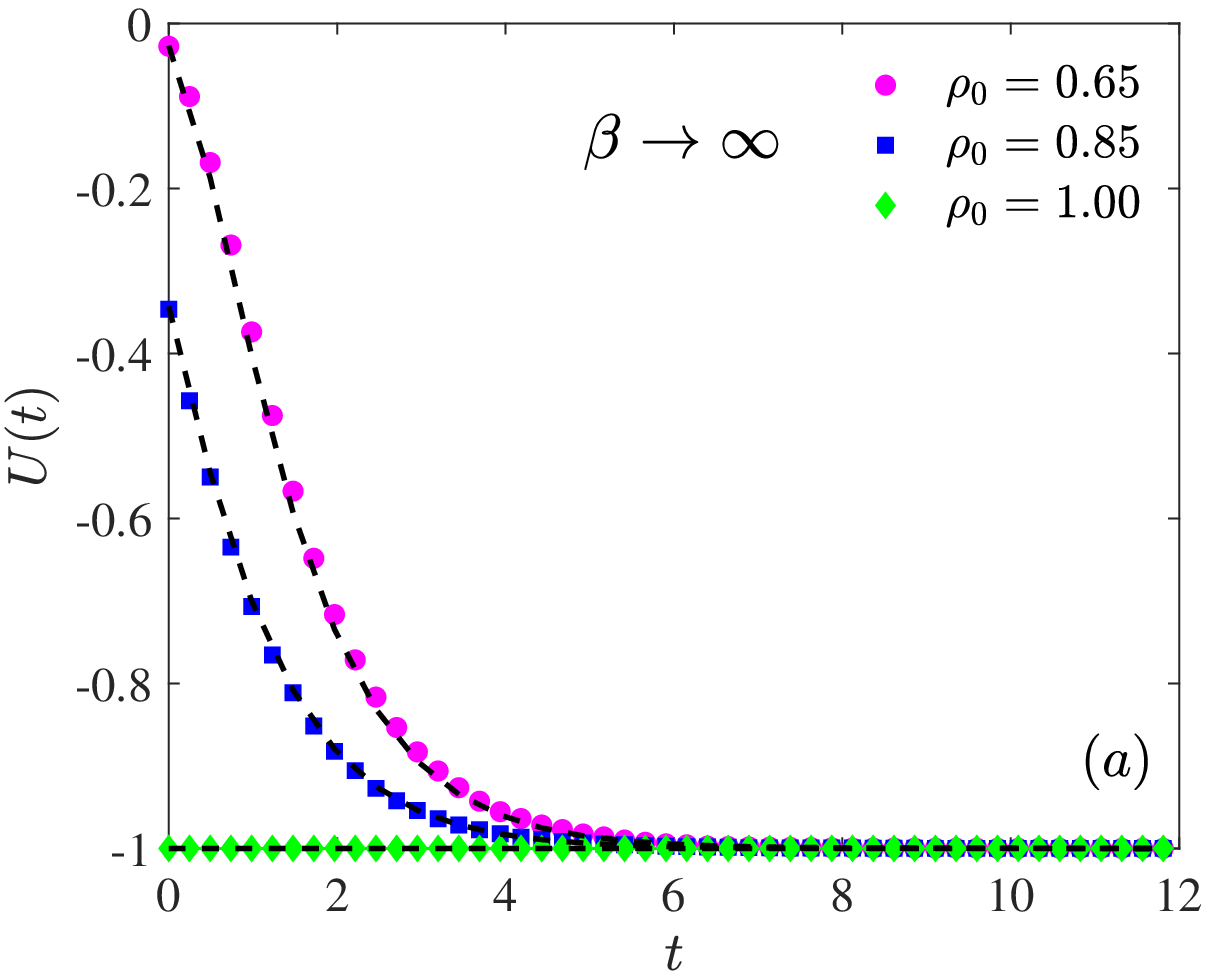}
\includegraphics[scale=.5]{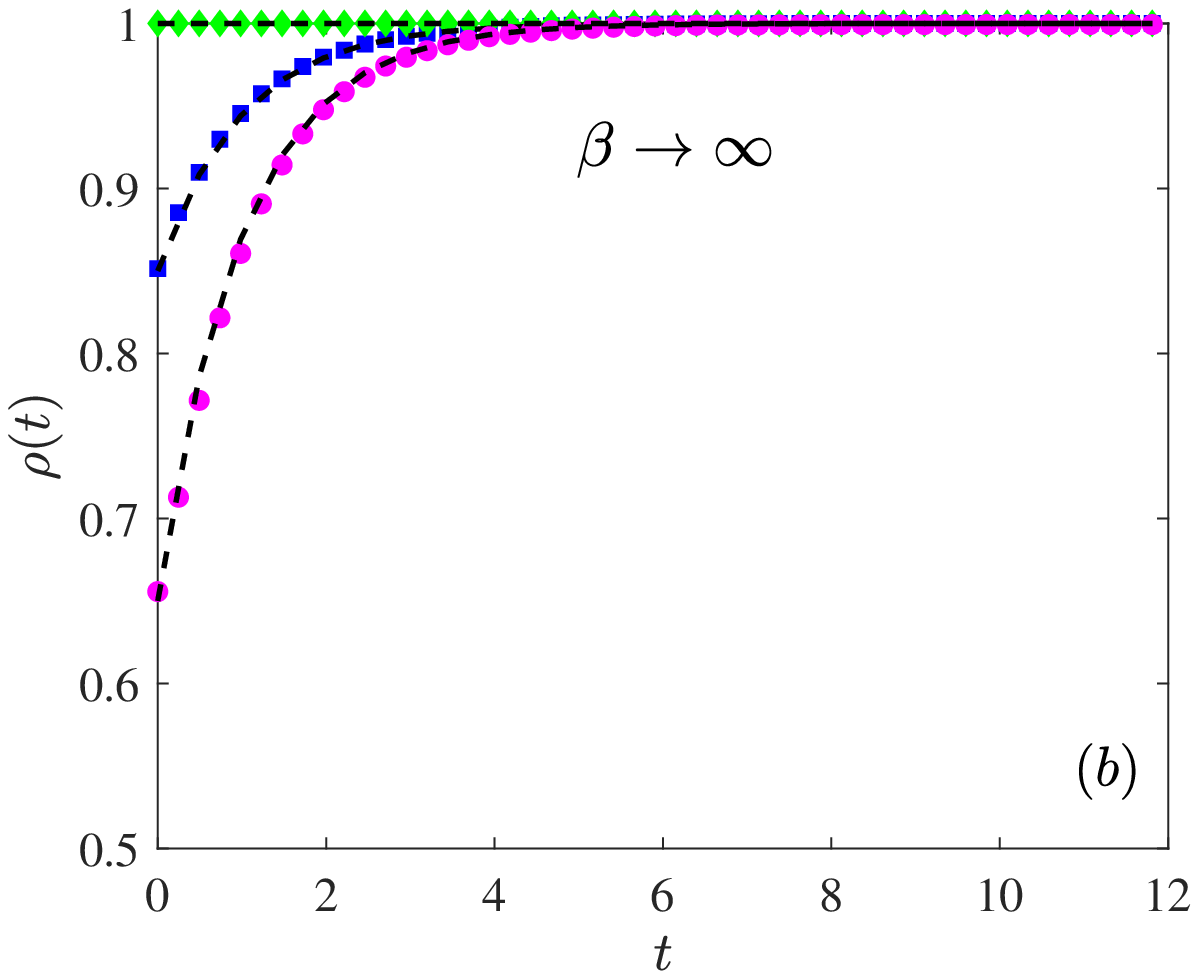}
\caption{The time evolution of (a) the total energy, $U$, and (b) the density of positive links, $\rho$, with different initial conditions of $\rho_0>1/2$ for $\beta\rightarrow\infty$. The dashed lines show our analytical solutions, i.e., Eqs. \ref{rho_t1} and \ref{U_t1}, which are in complete agreement with the simulations. The network size for all plots is $N=128$.}
\label{fig4}
\end{center}
\end{figure}

In this section, we simulate our proposed model, on networks with different sizes of $N$, and different initial conditions of $\rho_0$. Note also that we take each time step as $L$ simulation updates, so that at every time step, one link is updated, on average. At first, we focus on the fully random case with $\beta=0$. In Fig.~\ref{fig2}(a) and \ref{fig2}(b), we plot the time evolution of the energy, and the positive links' density, $\rho$, for different initial conditions of $\rho_0=0.0$, $0.2$, $0.5$, $0.8$, and $1.0$. The network size is $N=128$, here. For $\beta=0$, and from Eq. (\ref{p}), we have $p=1/2$, and thus, the updates are fully random. As it can be seen, at large $t$, the system approaches into a completely random configuration with $\rho_{\infty}=1/2$ and $U_{\infty}=0$. The network is imbalanced and has equal number of positive and negative links, as expected. Fig.~\ref{fig2}(c) shows the corresponding trajectory of the dynamics in $U-\rho$ space for two initial conditions of $\rho_0=0$ (with $U_0=+1$) and $\rho_0=1$ (with $U_0=-1$), both approaching into the final random state of $\rho=1/2$. Note that the dashed lines in (a), (b), and (c) show our analytical findings of Eqs. (\ref{rho_t}), (\ref{U_t}), and (\ref{U_rho}), respectively, which are all in complete agreement with our simulations.

\begin{figure}[ht]
\begin{center}
\includegraphics[scale=.5]{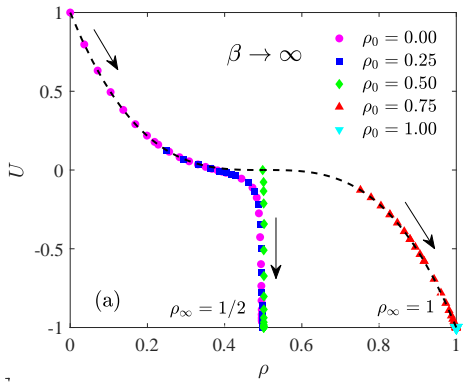}
\caption{The trajectory of the dynamics in $U-\rho$ space, for $\beta\rightarrow \infty$ with five different initial conditions. Note that representative points with $U=-1$ illustrate the balanced absorbing states. The dashed line represents our analytical solution of Eq.~\ref{U_rho} for fully uncorrelated trajectory. Note that as $\rho\rightarrow 1/2$, the deviation from the random trajectory increases. The network size is $N=128$.}
\label{fig5}
\end{center}
\end{figure}

To investigate the case of complete order, i.e., $\beta\rightarrow \infty$, Figs.~\ref{fig3} and \ref{fig4} show the time evolution of $U(t)$ and $\rho(t)$ for initial conditions of $\rho_0\leq 1/2$ and $\rho_0>1/2$, respectively. We see that the network always reaches a final balanced state of $U_\infty=-1$ at large $t$ (see Figs.~\ref{fig3}(a) and \ref{fig4}(a)). As it is observed in Fig.~\ref{fig3}(b), the final values of positive links' density, $\rho_\infty$, are independent of the initial conditions for all $\rho_0\leq 1/2$. Indeed, the network reaches a bipolar states with nearly same-size poles. However, for $\rho_0>1/2$, Fig.~\ref{fig4}(b) shows that $\rho_\infty \rightarrow 1$, and the final state is paradise. To better understand the evolution of the system, in Fig.~\ref{fig5}, we have also plotted $U$ versus $\rho$.  As it can be seen, the system finally approaches into one of its attractors of $\rho_\infty= 1/2$ or $\rho_\infty= 1$, depending on the corresponding initial conditions. Note here that in early stages of the dynamics, the evolution of the system coincides on the trajectory of an uncorrelated dynamics (see Fig.~\ref{fig2}(c)). The dashed lines in Figs. \ref{fig3} and \ref{fig4} show our analytical findings for large $\beta$. We observe that only for $\rho_0>1/2$, a good agreement between our analytic calculations and our simulations exists, and for $\rho_0\leq 1/2$, a large deviation occurs. In fact, analytic results show that the only balanced state is the paradise, and our simulations do not illustrate such a behavior for $\rho_0\leq 1/2$. We can understand this phenomenon qualitatively. At first, we note that as long as $\rho_0>1/2$, the conditions $\Delta U_{+-}>0$ and $\Delta U_{-+}<0$ holds, and thus Eq. (\ref{rho_t1}) is almost valid, and the density of positive links increases until it reaches unity where the stationary condition of $d\rho/dt=0$ holds. But, for the case of $\rho_0<1/2$, the above conditions also holds until $\rho\rightarrow 1/2$, at which $p^+\simeq p^-$, and thus $\rho$ is trapped into this value. It is worth to mention here that the reason that our rate equations cannot directly represent such a behavior is due to the deviation of the dynamics from its uncorrelated trajectory near $\rho=1/2$, which is observed in simulations. As depicted in Fig.~\ref{fig5}, for $\rho\rightarrow 1/2$, the deviation from the uncorrelated trajectory (dashed line) increases, which indicates that the mean-field approximation is not applicable here, and thus our analytical results are not able to cover the case of $\rho_0\leq1/2$.

\begin{figure}[ht]
\begin{center}
\includegraphics[scale=.5]{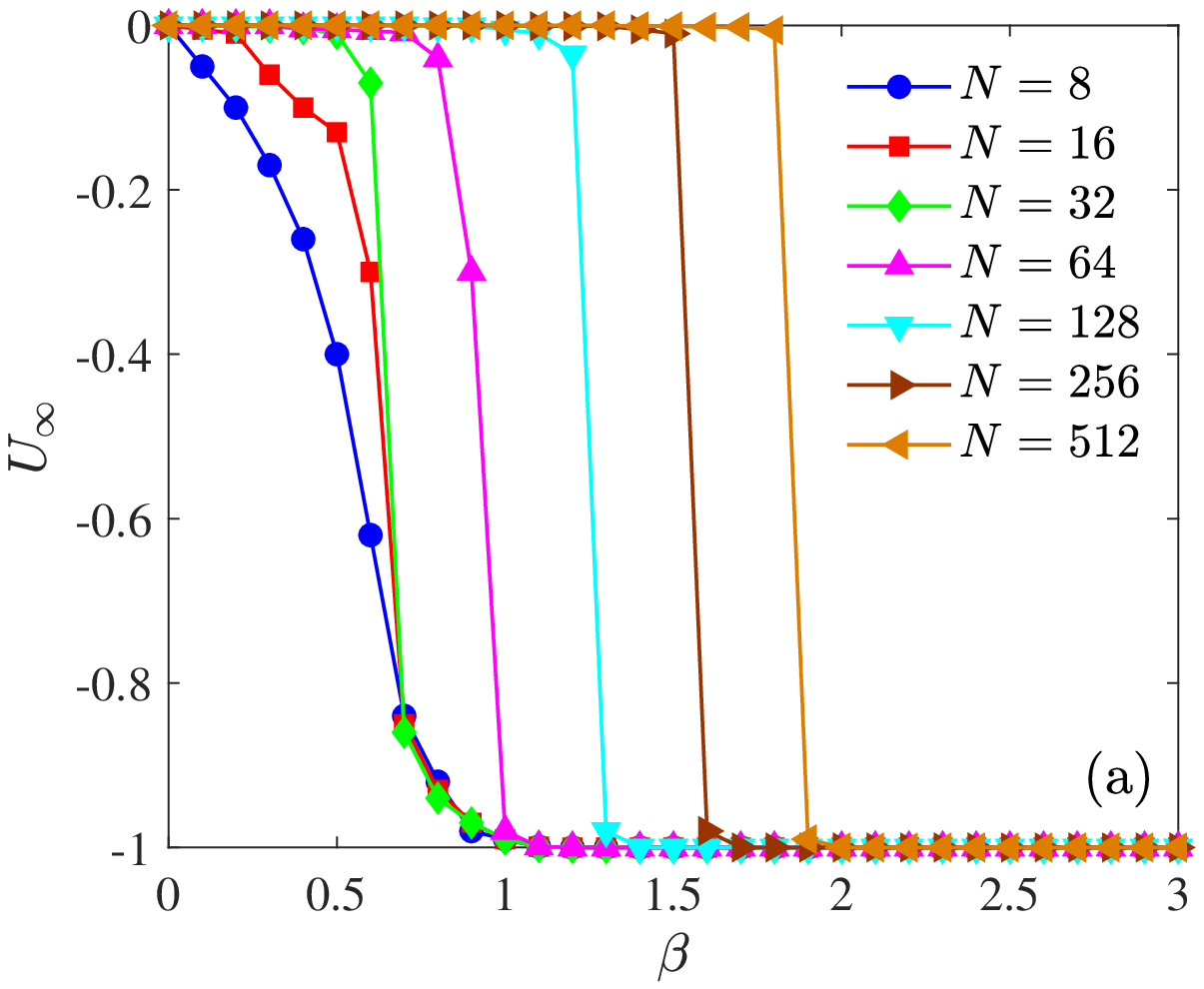}
\includegraphics[scale=.5]{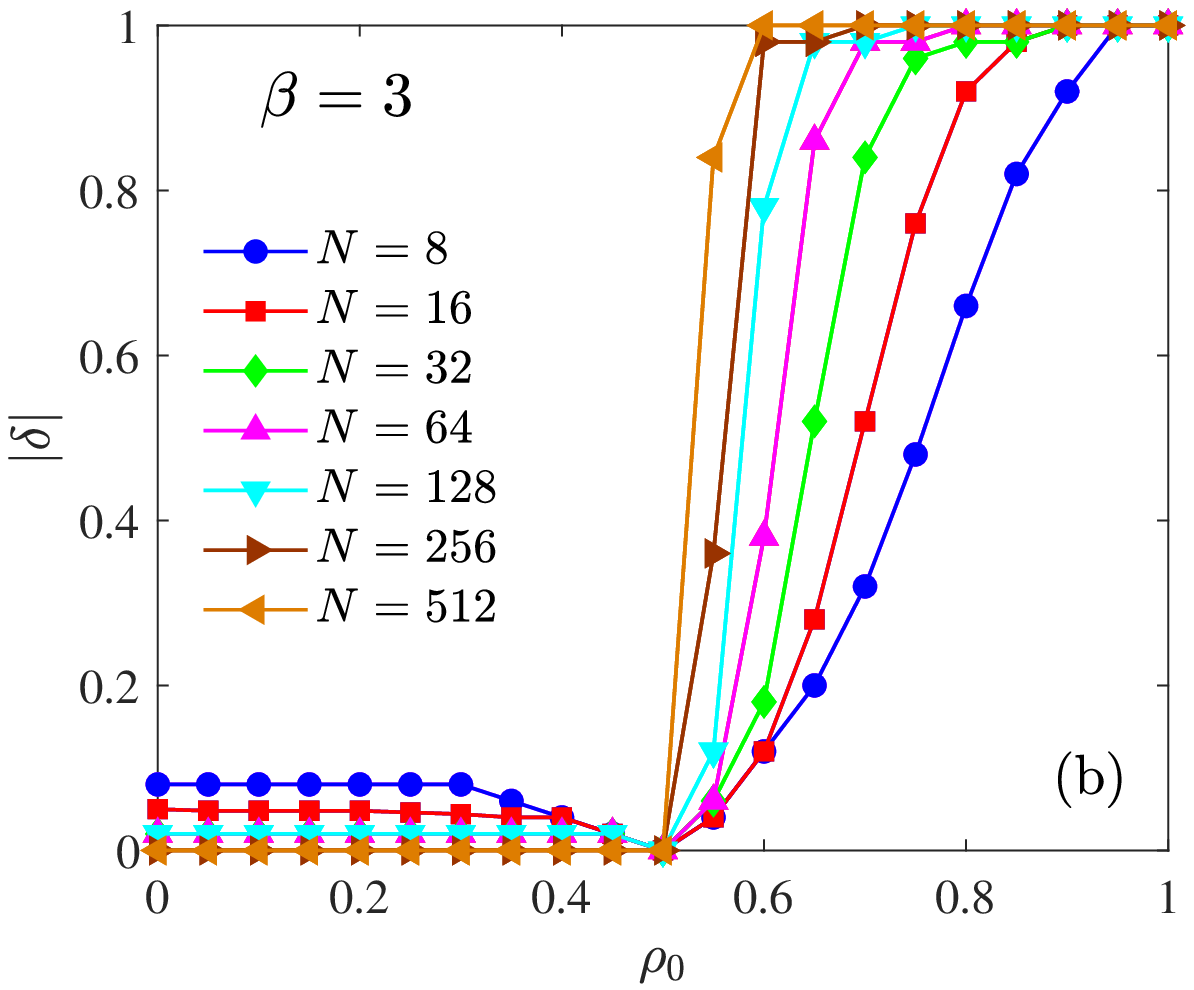}
\caption{(a) The $\beta$ dependency of large time behaviors of the total energy, $U$, for $\rho_0=0.6$. Different network sizes are shown with different symbols. For large networks, a nearly sharp transition occurs at some values of $\beta$, which goes to infinity as $N\rightarrow \infty$. (b) The order parameter, $\delta$, defined as the average link value, $\left\langle s\right\rangle$, as a function of $\rho_0$, for different network sizes of $N=8$, $16$, $32$, $64$, $128$, $256$, and $512$, at $\beta=3$. There is a sharp transition at $\rho_0=1/2$ for large networks.}
\label{fig6}
\end{center}
\end{figure}

Up to now, we have only discussed the limiting cases of $\beta=0$ and $\infty$. Clearly, a finite value of randomness $\beta$ is of more interest. For example, considering the fact that $\beta=1/kT$ and the fact that the dynamics of the links can be mapped to the spin dynamics in ferromagnetic (Ising) models, one can look for a possible phase transition at finite $\beta$. To study the intermediate randomness level, we calculate the behavior of final values of energy, $U_{\infty}$ for various $\beta$. Fig.~\ref{fig6}(a) shows plots of $U_\infty$ versus $\beta$ for $\rho_0=0.6$, for different network sizes. Interestingly, for large $N$, we find a nearly sharp transition from a random (imbalanced) state to an ordered (balanced) state, at $\beta=\beta_c$. In fact, for $\beta<\beta_c$, the dynamics takes the system into an imbalanced state with $U\neq -1$, and the positive link density satisfies the relation \ref{U_rho}, independent of the initial conditions. However, for $\beta\geq\beta_c$, the system finally becomes balanced, in the sense that $\rho_\infty\rightarrow 1/2$ and $1$ for $\rho_0\leq 1/2$ and $\rho_0>1/2$, respectively. As it can be seen, $\beta_c\rightarrow \infty$, as $N\rightarrow \infty$. It is important to note that our model leads to a final balanced state, despite large $\beta$, which is in contrary to the previous results where the system would be trapped in jammed states.

Further, it is interesting to check the effect of the network size on final values of the  order parameter in the balanced phase, i.e., the average link value, $\delta=\left\langle s_{ij}\right\rangle=2\rho_{\infty}-1$ as a function of $\rho_0$. In this respect, we plotted in Fig. \ref{fig6}(b), $|\delta|$ versus $\rho_0$ for a typical $\beta>\beta_c$ (here $\beta=3$). We find a sharp phase transition from bipolar state into paradise one, at $\rho_0=1/2$ for large system sizes. It also illustrates that only for finite networks we can have a bipolar state with $1/2<\rho_{\infty}<1$. It is worthmentioining here that we find a sharp transition at $\rho_0=1/2$, in comparison with the gradual transition at $\rho_0\simeq 0.65$, found in previous studies \cite{antal2005dynamics,antal2006social}.

\section{Conclusion}
\label{conc}

Balance theory has been used to study the solidarity and the stability of a social network. Much work has been done in studying the static and dynamic aspects of the balance theory. Almost all studied models take into account the assumption of tension reduction. However, social agents do not always change their relationship to reduce the tension. Also some previous models have shown that an unwanted outcome may emerge, where the system is trapped in an imbalanced state (a jammed state), forever. In this paper, we introduced a more realistic dynamical model by adding an intrinsic randomness, denoted by $1/\beta$, into the social agents behaviors. Our dynamics considers the possibility of both increasing and decreasing the tension, somehow that on average, the overall tension reduces. We find that it helps the system escape from the local minima in its energy landscape. Due to this feature, the dynamics is able to exit jammed states and finally is able to find a balanced state. This shows that our dynamics can solve the undesirable finding in previous studies. In addition, we find that for a critical value of the randomness, $1/\beta_c$, a transition occurs from an imbalanced into a balanced phase. Indeed, for $\beta>\beta_c$, the system approaches into two possible balanced states: bipolar or paradise. For finite networks, we showed that a final bipolar state with positive link density of $\rho_{\infty}>1/2$ can emerge. For large networks, a sharp phase transition for the size difference between two poles occurs at $\rho_0=1/2$. Indeed, by passing through this point, the system transitions from a bipolar state of $\rho_{\infty}=1/2$ into a paradise state of $\rho_{\infty}=1$. This \textit{sharp} transition can be compared with the \textit{gradual} transition in a nontrivial point of $\rho_0\simeq 0.65$, observed in previous models. 

We also derived an analytical description for the time evolution of the system, by applying a mean-field approximation. Our analytical solutions are in complete agreement with our simulations, except for the case of $\rho_0\leq 1/2$ for large $\beta$. We showed that this inconsistency is the result of deviation of the dynamics from the uncorrelated situation, as the system approaches $\rho=1/2$.

\section*{Acknowledgment}
PM would like to gratefully acknowledge the Persian Gulf University Research Council
for support of this work. AM also acknowledges support from Shiraz University Research Council.

\bibliographystyle{apsrev4-1}

\end{document}